\documentclass[twocolumn,smallcondensed,smallextended]{svjour3}

\usepackage{graphicx}
\usepackage{mathptmx}      
\usepackage{amsmath}
\usepackage{amsfonts}
\usepackage{bm}
\usepackage{dsfont}

\journalname{Quantum Information Processing}
\smartqed  

\newcommand{\openone}{\leavevmode\hbox{\normalsize1\kern-3.8pt\large1}}
\newcommand{\BEQ}{\begin{eqnarray}}
\newcommand{\EEQ}{\end{eqnarray}}
\newcommand{\ENQ}{\end{eqnarray}}
\newcommand{\BE}{\begin{eqnarray}}
\newcommand{\EE}{\end{eqnarray}}
\newcommand{\bq}{\begin{quote}}
\newcommand{\eq}{\end{quote}}
\newcommand{\nn}{\nonumber}

\newcommand{\inhoud}[2]{\hbox to #1{\hss #2 \hss}}
\newcommand{\forget}[1]{}
\newcommand{\ket}[1]{| \, #1 \rangle}

\newcommand{\bra}[1]{\langle \, #1  |}
\newcommand{\beq}{\begin{equation}}
\newcommand{\eeq}{\end{equation}}
\newcommand{\be}{\begin{equation}}
\newcommand{\ee}{\end{equation}}
\newcommand{\enq}{\end{equation}}

\newcommand{\1}{\mathds{1}}

\newcommand{\av}[1]{\langle #1 \rangle}
\newcommand{\Eq}[1]{(\ref{#1})}

\renewcommand{\H}{\mathcal{H}}

\begin{document}

\title{Monogamy of Correlations vs. Monogamy of Entanglement
}

\author{M.P. Seevinck}

\institute{M.P.~Seevinck \at - Institute for Theoretical Physics,\\
 Utrecht University, P.O.Box 80.195, 3508 TD  Utrecht,\\
- Centre for Time, Philosophy Department, University of Sydney, Main Quad A14, NSW 2006, Australia (visiting address)\\
\email{m.p.seevinck@uu.nl}
}

\date{Received: date / Accepted: date} 

\maketitle

\begin{abstract}
A fruitful way of studying physical theories is via the question whether the possible physical states and different kinds of correlations in each theory can be shared to different parties. Over the past few years it has become clear that both quantum entanglement and non-locality  (i.e., correlations that violate Bell-type inequalities) have limited shareability properties and can sometimes even be monogamous. We give a self-contained review of these results and present new results on the shareability of different kinds of correlations, including local,  quantum and no-signalling correlations. This includes an alternative simpler proof of the Toner-Verstraete monogamy inequality for quantum correlations, as well as a strengthening thereof. Further, the relationship between sharing non-local quantum correlations and sharing mixed entangled states is investigated, and already for the simplest case of bi-partite correlations and qubits this is shown to be non-trivial.  Also, a recently proposed new interpretation of Bell's theorem by Schumacher in terms of shareability of correlations is critically assessed. Finally, the relevance of monogamy of non-local correlations for secure quantum key distribution is pointed out, and in this regard it is stressed that not all non-local correlations are monogamous. 
\keywords{Quantum Mechanics \and Entanglement \and Non-locality \and Monogamy\and No-signalling \and Shareability \and Cryptography}
\PACS{ 02.50.Cw \and 03.67.Ud   \and 03.67.Mn \and 03.65.Ta}
\end{abstract}

\section{Introduction}
\label{intro}
It is more and more realised that entanglement is a physical resource \cite{horodeckis}. This has been the driving force behind the exploding field of quantum information theory, and has led to many operational and information-theoretic insights. More recently, and less well-known, it has been noted that non-locality (i.e., correlations that violate Bell-type inequalities) is also a resource for information theoretic tasks  \cite{barrett_key}.  
In this paper one aspect of their usefulness as a resource will be considered, namely that both entanglement and non-locality have limited shareability properties\footnote{Here `shareability' is meant kinematical and should not be thought of in a dynamical sense. But note that the kinematical shareability results are of course restrictive for any dynamics that one may wish to employ to actually describe a sharing process. Any such dynamics is bound by the constraints set by the kinematics; for the latter indicate what states of the desired kind exist at all in the theory's state space.}, and, in fact, can sometimes even be monogamous \cite{ckw,osborne,tonerverstraete,toner2}: consider three parties $a,b,c$ each holding a qubit, then if $a$'s and $b$'s qubits are maximally entangled, then $c$'s qubit must be completely unentangled to either $a$'s or $b$'s. Similarly if $a$ and $b$ are correlated in such a way that they violate the Clauser-Horne-Shimony-Holt (CHSH) inequality \cite{chsh} (this will also be called `non-locally correlated'), then neither $a$ nor $b$ be can be correlated in such a way (i.e., non-locally) to $c$ in any no-signalling theory.  
It has  been shown that such correlations can be used as a resource to distribute a secret key which is secure against eavesdroppers which are only constrained by the fact that any information accessible to them must be compatible with no-signalling, which is roughly the impossibility of arbitrarily fast signalling \cite{masanes_key}.
  
Classically none of this is possible since one does not have such monogamy trade-offs for states or correlations: all classical probability distributions can be shared \cite{toner2}. Indeed, if parties $a$, $b$ and $c$ have bits instead of quantum bits (qubits) and if $a$'s bit is perfectly correlated to $b$'s bit then there is no restriction on how $a$'s bit is correlated to $c$'s bit.  This difference in  shareability of states and in shareability of correlations is in fact one of the fundamental differences between classical and quantum physics, although it has only recently been properly studied. Fortunately, in the last few years we have been able to witness a number of fundamental results on both the shareability of quantum states and of correlations \cite{ckw,osborne,tonerverstraete,toner2,masanes06,koashi,barrett05}. 

The purpose of this paper is two-fold. Firstly, it intends to give a review of the recent results in both these fields (i.e., shareability of quantum states and of correlations), and, secondly, it adds new results to the latter, after which both fields are compared.  The review part of the paper draws heavily on work by Toner \cite{toner2} and Masanes, Acin \& Gisin \cite{masanes06}, but it is intended to be self-contained and tries to use a simple mathematical framework in terms of familiar mathematical objects of joint probability distributions for correlations. It uses the well-known CHSH inequality for expectation values of the product of local outcomes; it leaves aside the formulations in terms of information theoretic interactive proof systems and non-local games setups \cite{toner2,cleve}.  

More specifically, section \ref{shst} reviews the monogamy and shareability of entanglement, and section \ref{correlationsection} is devoted to  the monogamy and shareability of correlations. In subsection \ref{subsectionkindscorrelations} five different kinds of correlation are introduced whose shareability and monogamy aspects are reviewed in subsections \ref{sharecorr} and \ref{subsectionmonogamy}.  
Here we also prove that both unrestricted general correlations and so-called partially-local ones can be shared to any number of parties (called $\infty$-shareable). Next, subsection \ref{subsectionstronger} gives an alternative simpler proof of  the Toner-Verstraete monogamy inequality \cite{tonerverstraete} for quantum correlations, and this inequality is strengthened as well. Some of the results reviewed in this section have lead Schumacher \cite{schumacher} to argue for a new view on Bell's theorem, namely that it is a theorem about the shareability of correlations, and that, contrary to \emph{communis opinio}, its physical message does not at all deal with issues of locality or local realism. This argument will be presented and critically assessed in subsection \ref{interpretBell}. 
In section \ref{comparingmonogamies} we compare the results of sections  \ref{shst} and \ref{correlationsection}: we investigate the relationship between shareability of non-local quantum correlations and shareability of mixed entangled states, and already for the simplest case of bi-partite correlations this relationship will be shown to be non-trivial. For example, shareability of non-local correlations implies shareability of entanglement, but not vice versa.
Section \ref{crypto} addresses the possibilities  for cryptography of extracting a  secure secret key from correlations that are monogamous. It will be pointed out that some non-local correlations indeed suffice for this task, but that in general not all non-local correlations are monogamous (shown in section \ref{comparingmonogamies}) and that this fact should be crucially taken into account.
Finally, we will end with a short discussion in section \ref{discussion}.

\section{Shareability and monogamy of states}\label{shst}
\noindent
Let us first consider the shareability of states; that of correlations will be studied in the next section.  Classical states can be shared among many parties because one can just copy the state. Formally we can represented such a classical  copying procedure on a phase or configuration space where one uses the Cartesian product structure to relate systems and subsystems. That is, it is possible to extend any bi-partite pure state $S_{ab_1}=S_a \times S_{b_1}$ of the joint party $ab_1$ to $N-1$ other parties $b_2,\ldots b_N$ by considering the state $S_{a} \times S_{b_1} \times S_{b_2} \times \ldots \times S_{b_N}$, where $S_{b_1}=S_{b_i}, \forall i$. This ensures that the states $S_{ab_i}$ are identical to the original state $S_{ab_1}$. The bi-partite state $S_{ab_1}$ can thus be shared indefinitely. All this remains true under convex decompositions of pure states, and thus also for the case of mixed classical states. 

However, in quantum mechanics things are different. If a pure quantum state of two systems is entangled\footnote{A state $\rho$ of a bi-partite system $ab$ is entangled if and only if  it can not be written as a convex sum of product states: $\rho \neq \sum_i p_i \rho^a_i\otimes \rho^b_j$,  with $\sum_i p_i=1,0 \leq p_i \leq 1$ and $\rho^a_i$, $\rho^b_j$ states of the subsystems $a,b$ respectively. For the multipartite generalisation, which is not at all trivial, see \cite{seevuff08}.}, then none of the two systems can be entangled with a third system. This can be easily seen.  Suppose that systems\footnote{For ease of notation we will use the same symbols to refer to parties and the systems they possess, e.g., party $a$ possesses system $a$.} $a$ and $b$ are in a pure entangled state. Then when the system $ab$ is considered as part of a larger system, the reduced density operator for $ab$ must by assumption be a pure state. However, for the composite system $ab$ (or for any of its subsystems $a$ or $b$) to be entangled with another system, the reduced density operator of $ab$ must be a mixed state.  But since it is by assumption pure, no entanglement between $ab$ and any other system can exist. This feature is referred to as the monogamy of pure state entanglement\footnote{This is sometimes confusingly referred to as the claim that in quantum theory a system can be pure state entangled with only one other system \cite{spekkens}. But what about the GHZ state $(\ket{000}+\ket{111})/\sqrt{2}\,$? All three parties are entangled to each other in this pure state, so this seems to be a counterexample to the claim. What is actually meant is that if a pure state of two systems is entangled, then none of the two systems can be entangled with a third system. This is the formulation we will use.}. 

This monogamy can also be understood as a consequence of the linearity of quantum mechanics that is also responsible for the no-cloning theorem \cite{dieks82,wootters}. For suppose that party  $a$ has a qubit which is maximally pure state entangled to both a qubit held by party $b$ and a qubit held by party $c$. Party $a$ thus has a single qubit coupled to two perfect entangled quantum channels, which this party could exploit  to teleport two perfect copies of an unknown input state, thereby violating the no-cloning theorem, and thus the linearity of quantum mechanics \cite{terhal}.

If the state of two systems is not a pure entangled state but a mixed entangled state,  then it is possible that both of the two systems  are entangled to a third system. For example, the so-called $W$-state $\ket{\psi}=(\ket{001}+\ket{010}+\ket{100})/\sqrt{3}$ has bi-partite reduced states that are all identical and entangled. This feature is called `sharing of mixed state entanglement',  or `promiscuity of entanglement'. Entanglement is thus strictly speaking only monogamous in the case of pure entangled states. In the case of mixed entangled states  it can be promiscuous.  But this promiscuity is not unbounded:  although some entangled bi-partite states may be shareable with some finite number of parties, no entangled bi-partite state can be shared with an infinite number of parties\footnote{This is also referred to as `monogamy in an asymptotic sense'  by \cite{terhal}, but we believe that this feature is better captured by the term `no unbounded promiscuity'.}.
 Here  a bi-partite quantum state $\rho_{ab}$ is said to be $N$-shareable  when it is possible to find a quantum state $\rho_{a{b_1}{b_2}\ldots{b_N}}$ such that $\rho_{ab}=\rho_{{a{b_1}}}=\rho_{a{b_2}}=\ldots =\rho_{a{b_N}}$, where  $\rho_{a{b_k}}$ is the reduced state for parties $a$ and $b_k$. Consider the following theorem \cite{fannes,raggio}:  A bi-partite quantum state is $N$-shareable for all $N$ (also called $\infty$-shareable \cite{masanes06}) iff it is separable. Thus no bi-partite entangled state, pure or mixed, is $N$-shareable for all $N$.

The limited shareability of entanglement was first quantified by Coffman, Kundu \& Wootters \cite{ckw}. They gave  a trade-off relation
between how entangled $a$ is with $b$, and how entangled $a$ is with $c$ in a three-qubit system $abc$ that is in a pure state, using the measure of bi-partite entanglement called the tangle \cite{osborne}. It states that $\tau(\rho_{ab})+\tau(\rho_{ac})\leq \tau(\rho_{a(bc)})$ where $\tau(\rho_{ab})$ is the tangle\footnote{The tangle $\tau(\rho_{ab})$ is the square of the concurrence $C(\rho_{ab}):=\max\{0,\sqrt{\lambda_1}-\sqrt{\lambda_2}-\sqrt{\lambda_3}-\sqrt{\lambda_4}\}$, where the $\lambda_i$ are the eigenvalues of the matrix $\rho_{ab}(\sigma_y\otimes\sigma_y)\rho_{ab}^*(\sigma_y\otimes\sigma_y)$ in non-decreasing order, with $\sigma_y$ the Pauli-spin matrix for the $y$-direction.} between $a$ and $b$, analogous for $\tau(\rho_{ac})$ and $\tau(\rho_{a(bc)})$ is the bi-partite entanglement\footnote{In case of three qubits the tangle $\tau(\rho_{a(bc)})$ is equal to $4\,\textrm{det}\rho_a$, with $\rho_a=\textrm{Tr}_{bc}[\ket{\psi}\bra{\psi}]$ and $\ket{\psi}$ the pure three-qubit state.}
 across the bipartition $a$-$bc$. In general, $\tau$ can vary between $0$ and $1$, but monogamy constrains the entanglement (as quantified by $\tau$) that  party $a$ can have with each of parties $b$ and $c$. The generalisation to, possibly mixed, multi-qubit states has been recently proven by Osborne \& Verstraete \cite{osborne}:
 \begin{align}\label{EmonN}
 \tau(\rho_{ab_1}) +\tau(\rho_{ab_2})+\ldots+\tau(\rho_{ab_N})\leq \tau(\rho_{a(b_1b_2\ldots b_N)}).
 \end{align}
This is a general constraint on distributed entanglement and which quantifies the frustration of entanglement between different parties. For further investigations of the monogamy of entanglement, see also \cite{koashi,adesso,ou}.

\section{Shareability and monogamy of correlations}  \label{correlationsection}
\subsection{Kinds of correlations}\label{subsectionkindscorrelations}
We will review five different kinds of correlations that will be studied, as well as several useful mathematical characteristics of these correlations\footnote{This is a minimal review leaving out the discussion of the correlations in terms of convex sets and facets of polytopes. See \cite{barrett05,masanes06}, and chapter 2 in \cite{dissertation} for such  a more comprehensive overview.}.

\paragraph{General unrestricted correlations.}

Consider $N$ parties, labeled by $1,2,\ldots,N$, each holding a physical system that is to be measured using a finite set of different observables. Denote by $A_j$ the observable (random variable)  that party $j$ chooses (also called the setting $A_j$)  and by $a_j$ the corresponding measurement outcomes. We assume there to be only a finite number of discrete outcomes. 
The outcomes can be correlated in an arbitrary way. 
A general way of describing this situation, independent of the underlying physical model, is by a set of
joint probability distributions for the outcomes, conditioned on the settings chosen by the $N$ parties, where the correlations  are captured in terms of these joint probability distributions. They are denoted by 
\begin{align}\label{generalcorr}
P(a_1,\ldots,a_N|A_1,\ldots,A_N).
\end{align}
These probability distributions are assumed to be positive 
\begin{align}\label{posgen}
P(a_1,\ldots,a_N|A_1,\ldots,A_N)\geq 0,
\end{align}
and obey the normalization conditions
\begin{align}\label{norma}
\sum_{a_1,\ldots,a_N}P(a_1,\ldots,a_N|A_1,\ldots,A_N)=1.
\end{align}
We need not demand that the probabilities should not be greater than 1 because this follows from them being positive and from the normalization conditions.
We will now put further restrictions besides normalization on the probability distributions  (\ref{generalcorr})  that are motivated by physical considerations.

\paragraph{No-signalling correlations.}

 A no-signalling correlation  is a correlation 
  $P(a_1,\ldots,a_N|A_1,\ldots,A_N)$ such that one subset of parties, say parties $1,2,\ldots,k$, cannot signal to the other parties $k-1,\ldots,N$ by changing their measurement device settings $A_1,\ldots,A_k$. Mathematically this is expressed 
 as follows. The marginal probability distribution for each subset of parties only depends on the corresponding observables measured by the parties in the subset, e.g., for all outcomes $a_{1}$, $\ldots, a_{k}$: $P(a_1,\ldots, a_k|A_1,\ldots,A_N)=P(a_1,\ldots, a_k|A_1,\ldots,A_k)$ .   
 
  These conditions can all be derived from the following condition \cite{barrett05}. For each $k\in \{1,\ldots,N\}$ the marginal distribution that is obtained when tracing out $a_k$ is independent of what observable ($A_k$ or $A_k'$) is measured by party $k$:
  \begin{align}\label{nosignallingdistr}
  \sum_{a_k} P(a_1,\ldots,&a_k,\ldots, a_N|A_1,\ldots,A_k,\ldots,A_N)=\nonumber\\& \sum_{a_k} P(a_1,\ldots,a_k,\ldots, a_N|A_1,\ldots,A_k',\ldots,A_N),
  \end{align}
   for all outcomes $a_1,\ldots,a_{k-1},a_{k+1},\ldots,a_N$ and all settings $A_1,\ldots,A_k,A_k',\ldots A_N$.   This set of conditions ensures that all marginal probabilities are independent of the settings corresponding to the outcomes that are no longer considered. 
   In particular, \eqref{nosignallingdistr} the defines the marginal
   \begin{align}\label{marginalp}
   P(a_1,\ldots,a_{k-1},a_{k+1},\ldots, a_N|A_1,\ldots,A_{k-1},A_{k+1},\ldots,A_N),
   \end{align} for the $N-1$ parties not including party $k$. No-signalling ensures that it is not needed to specify whether $A_k$ or $A_k'$ is being measured by party $k$.

\paragraph{Local correlations.}

Local correlations are those that can be obtained if the parties are non-communicating and share classical information, i.e.,  they only have local operations and local hidden variables (also called shared randomness) as a resource. We take this to mean that these correlations can be written as
\begin{align}\label{localdistr}
P(a_1,\ldots, a_N|&A_1,\ldots,A_N)=\nn\\
&\int_\Lambda d\lambda p(\lambda) P(a_1|A_1,\lambda) \ldots P(a_N|A_N,\lambda),
\end{align}
where $\lambda\in\Lambda$ is the value of the shared local hidden variable, $\Lambda$ the space of all hidden variables and $p(\lambda)$ is the probability that a particular value of $\lambda$ occurs. Note that $p(\lambda)$ is independent of the outcomes $a_j$ and settings $A_j$, i.e., the settings are assumed to be `free variables'\cite{bell2}. Furthermore, $P(a_1|A_1,\lambda)$ is the probability that outcome $a_1$ is obtained by party $1$ given that the observable measured was $A_1$ and the shared hidden variable was $\lambda$, and similarly for the other terms $P(a_k|A_k,\lambda)$.  A correlation that is not local will be called non-local.

\paragraph{Partially-local correlations.}

Partially-local correlations are those  that can be obtained from an $N$-partite system in which subsets of the $N$ parties form extended  systems, whose internal states can be correlated in any way (e.g., signalling),
    which however behave local with respect to each other \cite{svetlichny,seevsvet}. Suppose provisionally that parties $1,\ldots,k$ form such a subset and the remaining parties $k+1,\ldots,N$ form another subset. The partially-local correlations can then be written as 
\begin{align}\label{partiallocaldistr}
P(a_1,\ldots,&\, a_N|A_1,\ldots,A_N)=\nn\\&\int_\Lambda d\lambda p(\lambda) P(a_1,\ldots,a_k|A_1,\ldots,A_k,\lambda) \times\nn\\&~~~~~~~~~~~~~~~~~~~P(a_{N-k},\ldots,a_N|A_{N-k},\ldots,A_N,\lambda),
\end{align}
 The  probabilities on the right hand side need not factorise any further. In case they would all fully factorise we retrieve the set of local correlations described above.

Formulas  similar to (\ref{partiallocaldistr}) with different partitions of the $N$-parties into two subsets, i.e., for different choices of the composing  parties and different values of $k$, describe other possibilities to give partially-local correlations. Convex combinations of these possibilities are also admissible.   We need not consider decomposition into more than two subsystems since any two subsystems in such  a decomposition can be considered jointly as parts of one subsystem still uncorrelated with respect to the others \cite{seevsvet}.   

\paragraph{Quantum correlations.}

Lastly, we consider the class of correlations that are obtained by general measurements on quantum states (i.e.,  those that can be generated if the parties share quantum states). These can be written as 
\begin{align}\label{quantumcorre}
P(a_1,\ldots, a_N|A_1,\ldots,A_N)=\textrm{Tr}[M_{a_1}^{A_1} \otimes\cdots\otimes M_{a_N}^{A_N} \rho].
\end{align}
Here $\rho$ is a quantum state (i.e., a unit trace semi-definite positive operator) on a Hilbert space $\H=\H_1\otimes\cdots\otimes\H_N$, where $\H_j$ is the quantum state space of the system held by party $j$. The sets $\{M_{a_1}^{A_1},\ldots,M_{a_N}^{A_N} \}$ define what is called a positive operator valued measure
 (POVM), i.e., a set of positive operators $\{ M_{a_j}^{A_j}\}$ satisfying 
$\sum_{a_j}M_{a_j}^{A_j}=\1,\forall A_j$.  
Of course, all operators $ M_{a_j}^{A_j}$ must commute for different $j$
in order for the joint probability distribution to be well defined, but this is ensured since for different $j$  the operators are defined for different subsystems (with each their own Hilbert space) and are therefore commuting. Note that \eqref{quantumcorre} is linear in both $M_{a_j}^{A_j}$ and $\rho$, which is  a crucial feature of quantum mechanics.

Quantum correlations are no-signalling and therefore the marginal probabilities derived from such correlations are defined in the same way as was done for no-signalling correlations (cf. Eq. \eqref{marginalp}).  For example, the marginal probability for party $1$ is given by $P(a_1|A_1)=\textrm{Tr}[M_{a_1}^{A_1}\rho^1]$, where $\rho^1$ is the reduced state for party 1.

\subsection{Shareability of correlations}\label{sharecorr}
General unrestricted correlations and local correlations can be shared. The latter fact is proven by  Masanes \emph{et al.} \cite{masanes06} and the first we will prove here. However, first we need the relevant definitions. Shareability of a general unrestricted probability distribution is defined as follows (where without loss of generality we restrict ourselves to shareability of bi-partite distributions). A bi-partite distribution $P(a,b_1|A,B_1,\ldots, B_N)$ is $N$-shareable with respect to the second party if an $(N+1)$-partite distribution $P(a,b_1,\ldots, b_{N}|A,B_1,\ldots, B_N)$ \mbox{exists} that is symmetric with respect to $(b_1,B_1), \,(b_2,B_2),\,\ldots,\,(b_N,B_N)$ and with marginals $P(a,b_i|A,B_1,\ldots, B_N)$ equal to the original distribution $P(a,b_1|A,B_1,\ldots, B_N)$, for all $i$. For notational clarity we use $b_i$ and $B_i$ (instead of $a_i$ and $A_i$) to denote outcomes and observables respectively for the parties other than the first party. If a distribution is shareable for all $N$ it is called  $\infty$-shareable \cite{masanes06} .

Shareability of a no-signalling probability distribution is defined analogously: A no-signalling $P(a,b_1|A,B_1)$  is $N$-shareable with respect to the second party if there exist an $(N+1)$-partite distribution $P(a,b_1,\ldots, b_{N}|A,B_1,\ldots, B_N)$ being symmetric with respect to $(b_1,B_1), \,(b_2,B_2),\,\ldots,\,(b_N,B_N)$ with marginals $P(a,b_i|A,B_i)$ equal to the original distribution $P(a,b_1|A,B_1)$, for all $i$. The difference between shareability of unrestricted correlations and of no-signalling correlations is that in the first case the marginals depend on all $N+1$ settings, whereas in the latter case they only depend on the two settings $A$ and $B_i$.

Suppose we are given a general unrestricted correlation $P(a,b_1|A,B_1,\ldots, B_N)$. We can then construct  
\begin{align} 
P(a,b_1,\ldots, b_{N}&|A,B_1,\ldots, B_N)=\nonumber\\
&P(a,b_1|A,B_1,\ldots, B_N)\delta_{b_1,b_2}\cdots\delta_{b_1,b_N},
\end{align}  
which has the same marginals $P(a,b_i|A,B_1,\ldots, B_N)$ equal to the original distribution $P(a,b_1|A,B_1,\ldots, B_N)$. This holds for all $i$, thereby proving the $\infty$-shareability. Thus an unrestricted correlation can be shared for all $N$.  If we restrict the distributions to be no-signalling, Masanes \emph{et al.} \cite{masanes06}  proved that if the distribution $P(a,b_1|A,B_1)$ is $N$-shareable then it satisfies all Bell-type inequalities with $N$ or less different settings $B_1$ (this extends a similar result  for quantum states by Terhal \emph{et al.} \cite{terhal03} and Werner \cite{werner}).

This result implies that there exists a local model of the form \eqref{localdistr} for correlations  $P(a,b|A,B)$  where  the first party has an arbitrary number and the second party has $N$ possible measurements  if  the correlations are $N$-shareable \cite{masanes06}. Indeed, suppose $P(a,b|A,B)$ is shareable to $N$ parties (labelled $B_i$, $i=1,\ldots,N$). The correlations between $A$ performed on party 1 and $B_i$ on party 2 are thus the same as the correlations between measurements of $A$ on party 1 and $B_i$ on the extra party $B_i$. Therefore,  the $N$ measurements $B_1,\ldots, B_N$ performed by party 2 can be viewed as one single joint measurement performed by $N$ parties  $B_i$ ($i=1,\ldots,N$), and it is known that there always exists a local model when all but one of the parties perform only one measurement.
In particular, this implies that two-shareable states can not violate the CHSH inequality; see also section \ref{interpretBell} that assesses the foundational relevance of this result.  Furthermore,  Masanes \emph{et al.} \cite{masanes06}  proved that $\infty$-shareability implies that the $N$-partite distribution is local (in the sense of Eq. \eqref{localdistr}) for all $N$ and that the converse holds as well. Local correlations can thus be shared indefinitely, and vice versa.

\subsection{Monogamy of correlations}\label{subsectionmonogamy}
Because general unrestricted correlations and local ones can be shared indefinitely they both will not show any monogamy constraints. This implies that partially-local correlations (see Eq. \eqref{partiallocaldistr}) also do not show any monogamy, since these are combinations of local and general unrestricted correlations between subsystems of the $N$-systems. However, and perhaps surprisingly, quantum and no-signalling correlations are not $\infty$-shareable and they must therefore show monogamy constraints, as will now see \footnote{Intuitively this can be understood as follows. Because the class of general unrestricted correlations is richer than the class of no-signalling correlations, and in fact contains all possible correlations, there is no possibility of leaving this class of correlations when sharing them.  To the contrary, the class of no-signalling correlations is rather limited; indeed, it turns out that not all such correlations can be shared while requiring that one stays in this class. When trying to construct the shared no-signalling states one will end up with resulting states that are signalling rather than no-signalling.}.
First, consider a very strong monogamy property for extremal no-signalling correlations, mentioned by Barrett \emph{et al.} \cite{barrett05}. Suppose one has some no-signalling three-party probability distribution $P(a,b,c|A,B,C)$ for parties $a$, $b$ and $c$. In case the marginal distribution\\ $P(a,b|A,B)$ of system $ab$ is an extremal no-signalling correlation\footnote{A no-signalling correlation is extremal iff it is a vertex of the bi-partite no-signalling polytope which means that any convex decomposition in terms of no-signalling correlations is unique.} then it cannot be correlated to the third system $c$:
\begin{align}
P(a,b,c|A,B,C)=P(a,b|A,B)P(c|C),
\end{align}
in other words,  the extremal correlation $P(a,b|A,B)$ is completely monogamous.  Barrett \emph{et al.} \cite{barrett05} show that this implies that all Bell-type inequalities for which the maximal violation consistent with no-signalling is attained by a unique correlation have monogamy constraints.  
An example is the CHSH inequality, as will be shown below.

Extremal no-signalling correlations thus show monogamy, but what about non-extremal no-signalling correlations?  Just as was the case for quantum states where non-extremal (mixed state) entanglement can be shared (See Eq. \eqref{EmonN}), non-extremal no-signalling correlations can be shared as well.  This can be shown in terms of the well known Bell-type experimental setup where each of the two parties $a$ and $b$ implements two possible dichotomous observables, $A,A'$ and $B,B'$ respectively.  The CHSH inequality $|\av{\mathcal{B}_{ab}}|\leq2$
is the only non-trivial local Bell-type inequality for this setup \cite{chsh}. Here $\mathcal{B}_{ab}=AB+AB'+A'B-A'B'$ is called the CHSH polynomial (or CHSH operator in the quantum case) for the bi-partite system $ab$. 

No-signalling correlations obey the following tight trade-off relation in terms of the CHSH operators $\mathcal{B}_{ab}$ and $\mathcal{B}_{ac}$ for party $ab$ and $ac$ respectively, as first proven by Toner \cite{toner2}: 
\begin{align}\label{monogamynosignalling}
|\av{\mathcal{B}_{ab}}_{\textrm{ns}}| +|\av{\mathcal{B}_{ac}}_{\textrm{ns}}|\leq 4.
\end{align} 
Here $\mathcal{B}_{ac}=AC+AC'+A'C-A'C'$ is the CHSH polynomial for parties $a$ and $c$, and $\av{\mathcal{B}_{ab}}_{\textrm{ns}}$ is the expectation value\footnote{The subscript `ns' indicates that the expectation value is for no-signalling correlations \eqref{nosignallingdistr}.} of the CHSH operator $\mathcal{B}_{ab}$  for a no-signalling correlation \eqref{nosignallingdistr}, and analogous for $\av{\mathcal{B}_{ac}}_{\textrm{ns}}$. Tightness was shown by Toner \cite{toner2}: for any pair $\av{\mathcal{B}_{ab}}_{\textrm{ns}}$, $\av{\mathcal{B}_{ac}}_{\textrm{ns}}$ that obeys  \eqref{monogamynosignalling} there is a no-signalling correlation with these expectation values. A particular multipartite generalisation of \eqref{monogamynosignalling} for a large class of linear multi-partite Bell-type inequalities 
has been recently achieved by Paw\l owski  \& Brukner \cite{pawlowskibrukner}.

\begin{figure}[!hb]
\includegraphics[scale=1.2]{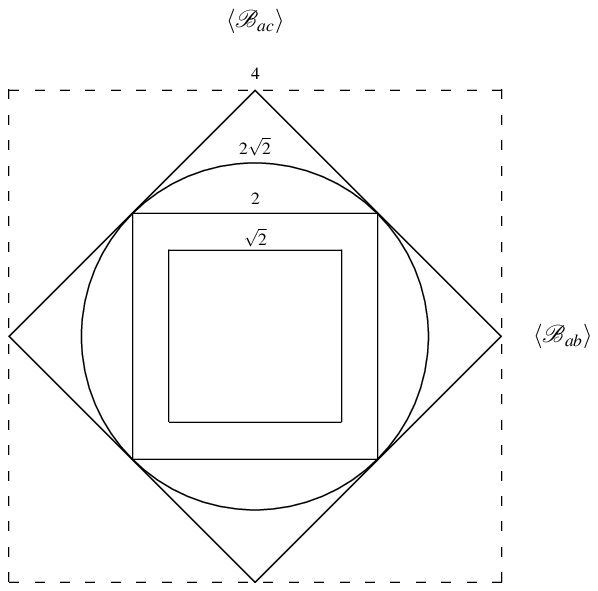}
\caption{The space $\av{\mathcal{B}_{ab}}$-$\av{\mathcal{B}_{ac}}$ of allowable values for the CHSH operators for systems $ab$ and $ac$.  General unrestricted correlations can reach the absolute maximum which is the largest square with edge points $(\pm4,\pm4)$.  All quantum correlations lie within the circle, and all no-signalling correlations lie within the tilted square. For comparison the local correlations are also shown. These lie within the square with edge points $(\pm2,\pm2)$. The correlations obtainable by orthogonal measurements on separable two-qubit states lie within the smallest square with edge points $(\pm\sqrt{2},\pm\sqrt{2})$. Figure adapted from \cite{tonerverstraete}.}\label{figmonogamy}
\end{figure}

This trade-off relation is  depicted in the tilted square of Figure \ref{figmonogamy}.  Extremal no-signalling correlations can attain $|\av{\mathcal{B}_{ab}}_{\textrm{ns}}|=4$, but then necessarily $|\av{\mathcal{B}_{ac}}_{\textrm{ns}}|=0$, and vice versa (this is monogamy of extremal no-signalling correlations), whereas non-extremal ones are shareable since the correlation terms $|\av{\mathcal{B}_{ab}}_{\textrm{ns}}|$ and  $|\av{\mathcal{B}_{ac}}_{\textrm{ns}}|$ can both be non-zero at the same time. But note that in case the no-signalling correlations are non-local they can not be shared, i.e., it is not possible that 
$|\av{\mathcal{B}_{ab}}_{\textrm{ns}}|\geq2$ and also $|\av{\mathcal{B}_{ac}}_{\textrm{ns}}|\geq 2$ (a fact already shown in  \cite{masanes06}). This shows that if these non-local correlations can be shared they must be signalling.  Alternatively, this can also be phrased as follows. In order to be non-local with both party $b$ and $c$, and also remain no-signalling, party $a$ is faced with an unsolvable dilemma in choosing her measurements, which would need to be different  in $\mathcal{B}_{ab}$ and  $\mathcal{B}_{ac}$ 
This feature is termed `monogamy of non-local correlations'. As a corollary it follows that if $N+1$ parties $a,b_1,b_2,\ldots,b_N$ share some correlations (e.g., via a quantum state) and each chooses to measure one of two observables, then $a$ violates the CHSH inequality with at most one party $b_i$.

We have seen that for general unrestricted correlations no monogamy constraints hold. They can thus reach the largest square in Figure \ref{figmonogamy}, i.e., $|\av{\mathcal{B}_{ab}}|$ and $|\av{\mathcal{B}_{ac}}|$ are not mutually constrained and can each obtain a value of $4$, so as to give the absolute maximum of the left hand side of (\ref{monogamynosignalling}) which is the value $8$. The monogamy bound (\ref{monogamynosignalling}) therefore gives a way of discriminating no-signalling from general correlations:  if it is violated the correlations must be signalling. These lie outside the tilted square but inside the largest square in Figure \ref{figmonogamy}.

For local correlations no such trade-off as in Eq. \eqref{monogamynosignalling} or (\ref{tonerverstraeteineq}) holds. Indeed, it is possible to have\footnote{Here $\av{\mathcal{B}_{ab}}_{\textrm{local}}$ is the expectation value of the CHSH operator for local correlations \eqref{localdistr} between party $a$ and $b$.} both $|\av{\mathcal{B}_{ab}}_{\textrm{local}}|=2$ and $|\av{\mathcal{B}_{ac}}_{\textrm{local}}|=2$, see also Figure \ref{figmonogamy}. This reflects the fact that local correlations are always shareable.

Let us finally consider correlations that result from making local measurements on quantum systems.   All quantum correlations that violate the CHSH inequality are monogamous as follows from the following tight trade-off inequality for a three-partite system $abc$ proven by Toner \& Verstraete \cite{tonerverstraete}\footnote{It should be noted that early results in this direction have been obtained by Krenn \& Svozil \cite{krenn} (Section 5) and by Scarani \& Gisin \cite{scarani} (Theorem 1).}:
\begin{align}
\av{\mathcal{B}_{ab}}_{\textrm{qm}}^2 +\av{\mathcal{B}_{ac}}_{\textrm{qm}}^2\leq8,
\label{tonerverstraeteineq}
\end{align}  
where $\mathcal{B}_{ab}$ is the CHSH operator for parties $a$ and $b$, and analogous for $\mathcal{B}_{ac}$ (here, and in the following,  the subscript 'qm' denotes that the expectation value is determined for quantum correlations \eqref{quantumcorre}). The correlations admissible by this trade-off relation lie in the interior of the circle in Figure \ref{figmonogamy}. Toner and Verstraete \cite{tonerverstraete} have explicitly shown tightness. Thus the inequality \eqref{tonerverstraeteineq} gives exactly the allowed values of $(\av{\mathcal{B}_{ab}}_{\textrm{qm}}$, $\av{\mathcal{B}_{ac}}_{\textrm{qm}})$. Note that as a corollary the Tsirelson inequality \cite{cirelson} follows: $|\av{\mathcal{B}_{ab}}_{\textrm{qm}}|$, $\,|\av{\mathcal{B}_{ac}}_{\textrm{qm}}|\leq2\sqrt{2}$.

Just as was the case for no-signalling correlations, quantum correlations show an interesting trade-off relationship: In case the quantum correlations between party $a$ and $b$ are non-local (i.e., when $|\av{\mathcal{B}_{ab}}_{\textrm{qm}}|>2$) the correlations between parties $a$ and $c$ cannot be non-local (i.e., necessarily $|\av{\mathcal{B}_{ac}}_{\textrm{qm}}|\leq 2$), and vice versa (cf. \cite{scarani}). These non-local quantum correlations can thus not be shared. Furthermore, in case they are maximally non-local, i.e., $|\av{\mathcal{B}_{ab}}_{\textrm{qm}}|=2\sqrt{2}$ the other must be uncorrelated, i.e., it must be that   $|\av{\mathcal{B}_{ac}}_{\textrm{qm}}|=0$, and vice versa.

Note that if $ab$ and $ac$ each share a maximally entangled pair, there are sets of measurements such that either 
$\av{\mathcal{B}_{ab}}_{\textrm{qm}}$ or $\av{\mathcal{B}_{ac}}_{\textrm{qm}}$ is $2\sqrt{2}$. 
But, as noted by Toner \cite{toner2}, this does not contradict \eqref{tonerverstraeteineq}: it is required that the measurements performed by party $a$ are the same in both $\mathcal{B}_{ab}$ and in $\mathcal{B}_{ac}$.  This is analogous to the requirement that was needed in section \ref{shst} to show monogamy of entanglement, namely that $b$ and $c$ are entangled with the same qubit of $a$.

The correlations  that separable quantum states allow for are shareable. Indeed, in the $\av{\mathcal{B}_{ab}}_{\textrm{qm}}$-$\av{\mathcal{B}_{ac}}_{\textrm{qm}}$ plane of Figure \ref{figmonogamy} such correlations can reach the full square with edge length $2$. However, when considering qubits and measurements that are restricted to orthogonal ones only (e.g., Pauli spin observables  $\sigma_x,\sigma_y, \sigma_z$ in the $x,y,z$-directions) one obtains tighter bounds; see  \cite{uffseev}. In such a case  the possible values 
 are restricted to the smallest square of Figure \ref{figmonogamy}:  $|\av{\mathcal{B}_{ab}}_{\textrm{qm}}|,$ $|\av{\mathcal{B}_{ac}}_{\textrm{qm}}|\leq \sqrt{2}$. But again there is no trade-off on the shareability of the correlations in separable states since this full square can be reached.

\subsection{A stronger monogamy relation for non-local quantum correlations}\label{subsectionstronger}

We will now give an alternative simpler proof of  the inequality (\ref{tonerverstraeteineq}) than was given by Toner \& Verstraete \cite{tonerverstraete}, and one that also allows us to strengthen this result  as well.
The proof uses the idea that (\ref{tonerverstraeteineq}), which describes the interior of a circle in the  $\av{\mathcal{B}_{ab}}$-$\av{\mathcal{B}_{ac}}$ plane,  is equivalent to the interior of the set of tangents to this circle. It is thus 
a compact way of writing the following infinite set of linear equalities
\begin{align}\label{linmon}
\mathcal{S}=\max_\theta \av{\mathcal{S}_\theta}_{\textrm{qm}}\leq 2\sqrt{2},
\end{align}
where we have used $\sqrt{x^2 +y^2} =\max_\theta (\cos\theta\, x +\sin\theta\, y)$, and where 
$\mathcal{S}_{\theta}=\cos\theta\, \mathcal{B}_{ab}+\sin\theta \,\mathcal{B}_{ac}$. 

We will now prove this by showing that $|\av{\mathcal{B}_{ab} \cos\theta + \mathcal{B}_{ac} \sin\theta }_{\textrm{qm}}|\leq 2\sqrt{2}$ for all $\theta$, using  a method presented by \cite{dieks} in a different context.  In this proof we only consider quantum correlations,  so for brevity we drop the subscript `qm' from the expectation values. Let us first write 
\begin{align}
\mathcal{B}_{ab} \cos\theta + \mathcal{B}_{ac} &\sin\theta =(A+A')B\cos\theta +(A-A')B'\cos\theta\nonumber\\&+ (A+A')C\sin\theta +(A-A')C\sin\theta.
\label{tussen}
\end{align}
Next we use the fact that in this context it is sufficient to consider qubits and projective measurements that are real and traceless only \cite{masanes,tonerverstraete}.  Let us thus express $A$ and $A'$  in terms of orthogonal Pauli observables $\sigma_z,\sigma_x$ for measurements in the $z$ and $x$ direction respectively: $A= \cos\gamma \sigma_x +\sin\gamma\sigma_z$ and $A'= \cos\gamma \sigma_x -\sin\gamma\sigma_z$. This gives $A+A'=2\cos\gamma\sigma_x,~ A-A'=2\sin\gamma \sigma_z$.
%
%
%
%
%
%
Taking the expectation value of (\ref{tussen}) gives
\begin{align}
|\av{\mathcal{B}_{ab} \cos\theta}_{ab} &+ \av{\mathcal{B}_{ac} \sin\theta}_{ac}|= \\&2| \av{\sigma_x B}_{ab}\cos\gamma\cos\theta +\av{\sigma_z B'}_{ab}\sin\gamma\cos\theta\nn\\\nn& ~~+\av{\sigma_x C}_{ac}\cos\gamma\sin\theta  +\av{\sigma_z C'}_{ac}\sin\gamma\sin\theta |
\end{align}
The right hand side can be considered to be twice the absolute value of the inproduct of the two four-dimensional vectors $\bm{a}=(\av{\sigma_x B}_{ab}, \av{\sigma_z B'}_{ab},\av{\sigma_x C}_{ac},\av{\sigma_z C'}_{ac})$ and $\bm{b}=
(\cos\gamma\cos\theta,$ $ \sin\gamma\cos\theta,$$ \cos\gamma\sin\theta,$$\sin\gamma\sin\theta)$. If we now apply the Cauchy-Schwartz  inequality  $|(\bm{a},\bm{b})|\leq ||\bm{a}||\,||\bm{b}||$ we find, for all $\theta$: 
\begin{align}
|\av{\mathcal{B}_{ab}& \cos\theta}_{ab} + \av{\mathcal{B}_{ac} \sin\theta}_{ac}|\nn\\
&\leq2\sqrt{\av{\sigma_xB}_{ab}^2+\av{\sigma_zB'}_{ab}^2+ \av{\sigma_xC}_{ac}^2+\av{\sigma_zC'}_{ac}^2}~\times\nn\\&\qquad
\sqrt{\cos^2\gamma(\cos^2\theta+ \sin^2\theta)+\sin^2\gamma(\cos^2\theta+ \sin^2\theta)}\nn\\
&\leq 2\sqrt{2(\av{\sigma_x}_{a}^2+\av{\sigma_z}_{a}^2)}\nn\\
&\leq2 \sqrt{2}\sqrt{1-\av{\sigma_y}_{a}^2}\label{proof1}\\
&\leq 2\sqrt{2}.\label{proof2}
\end{align}
This proves (\ref{linmon}). Here we have used that $\av{\sigma_x}_{\textrm{qm}}^2+\av{\sigma_y}_{\textrm{qm}}^2 +\av{\sigma_z}_{\textrm{qm}}^2\leq1$ for all single qubit quantum states, and for clarity we have used the subscripts $ab$, $ac$ and $a$ to indicate with respect to which subsystems the quantum expectation values are taken. 
Using (\ref{proof1}) we obtain 
\begin{align}
\av{\mathcal{B}_{ab}}_{\textrm{qm}}^2 +\av{\mathcal{B}_{ac}}_{\textrm{qm}}^2\leq8(1-\av{\sigma_y}_{a}^2),
\label{strongmon}
\end{align} which strengthens the original monogamy trade-off inequality (\ref{tonerverstraeteineq}). An alternative, but similar strengthening  of  (\ref{tonerverstraeteineq}) 
was already found in \cite{tonerverstraete}: $\av{\mathcal{B}_{ab}}_{\textrm{qm}}^2 +\av{\mathcal{B}_{ac}}_{\textrm{qm}}^2\leq8(1-\av{\sigma_y\sigma_y}_{bc}^2).$

So far we have only focused on subsystems $ab$ and $ac$, and not on the subsystem $bc$. One could thus also consider the quantity 
$\av{\mathcal{B}_{bc}}_{\textrm{qm}}$.  The above method would give the intersection of the three cylinders $\av{\mathcal{B}_{ab}}_{\textrm{qm}}^2+\av{\mathcal{B}_{ac}}_{\textrm{qm}}^2\leq8$, $\av{\mathcal{B}_{ab}}_{\textrm{qm}}^2+\av{\mathcal{B}_{bc}}_{\textrm{qm}}^2\leq8$, $\av{\mathcal{B}_{ac}}_{\textrm{qm}}^2+\av{\mathcal{B}_{bc}}_{\textrm{qm}}^2\leq8$. But it is known \cite{tonerverstraete} that this bound is not tight.

It might be tempting to think that because of these results we could have the following inequality, which is even stronger than  (\ref{tonerverstraeteineq}):
\beq\label{mono1}
\av{\mathcal{B}_{ab}}_{\textrm{qm}}^2+\av{\mathcal{B}_{ac}}_{\textrm{qm}}^2+\av{\mathcal{B}_{bc}}_{\textrm{qm}}^2\leq8.
\eeq
However, this is not true. For a pure separable state (e.g., $\ket{000}$) the left hand side has a maximum of $12$, which violates \Eq{mono1}. But inequality (\ref{mono1}) is true for the exceptional case that we have maximal violation for one pair, say $ab$, since we know from (\ref{tonerverstraeteineq}) that $\av{\mathcal{B}_{ac}}_{\textrm{qm}}$ and  $\av{\mathcal{B}_{bc}}_{\textrm{qm}}$ for the  other two pairs must then be zero.
We can see the monogamy trade-off at work: in case of maximal violation of the CHSH inequality (i.e., for maximal entanglement) the left hand side of \Eq{mono1} has a maximum of 8, whereas in case of no violation of the CHSH inequality it allows for a maximum value of 12, which can be obtained by pure separable states. Thus we see the opposite behavior from what is happening in the ordinary CHSH inequality: for the expression considered here (i.e., the left-hand side of \eqref{mono1}),  separability gives higher values, and entanglement necessarily lower values.

A correct bound is obtained from (\ref{proof1}) and the two similar ones for the other two expressions $\av{\mathcal{B}_{ab}}_{\textrm{qm}}^2+\av{\mathcal{B}_{bc}}_{\textrm{qm}}^2$ and  $\av{\mathcal{B}_{ac}}_{\textrm{qm}}^2+\av{\mathcal{B}_{bc}}_{\textrm{qm}}^2$. This gives:
\begin{align}
\av{\mathcal{B}_{ab}}_{\textrm{qm}}^2+\av{\mathcal{B}_{ac}}_{\textrm{qm}}^2+&\av{\mathcal{B}_{bc}}_{\textrm{qm}}^2\leq\nn\\
&12- 4(\av{\sigma_y}_a^2+\av{\sigma_y}_b^2+\av{\sigma_y}_c^2).
\end{align}
 However, it is unknown if this inequality is tight.

\subsection{Interpreting Bell's theorem}\label{interpretBell}

In subsection \ref{sharecorr} it was pointed out that non-local correlations, either quantum or no-signalling, can be completely monogamous, whereas $\infty$-shareability and locality of correlations are equivalent properties. Furthermore, it was shown that  there exists a local model of the form \eqref{localdistr} for correlations  $P(a,b|A,B)$  when the first party has an arbitrary number and the second party has $N$ possible measurements  if and only if the correlations are $N$-shareable. 

This has led Schumacher \cite{schumacher} to argue for a new view on Bell's theorem, which, according to \emph{communis opinio},  states that quantum mechanics is non-local\footnote{To be more specific: Bell's theorem states that quantum correlations exist that cannot be reproduced in terms of local correlations of the form \eqref{localdistr}.}: it is a theorem about the shareability of correlations, and its physical message is not at all about issues of locality or local realism\footnote{For an outline of the doctrine of local realism see \cite{bell2} or chapter 2 of \cite{dissertation}.}. Schumacher argues that 2-shareability of correlations is sufficient to get a conflict with quantum mechanics because it implies the CHSH inequality from which we already know that such a conflict follows. He also stresses that the assumption of 2-shareability of correlations  is a weaker assumption than the assumption of full-blown local realism, since the latter implies $\infty$-shareability.  From this he concludes that the real physical message of Bell's theorem is that quantum mechanical correlations are in general not 2-shareable; and \emph{not} that quantum mechanics is non-local in some way or another.

\begin{figure}[h]
\includegraphics[scale=0.85]{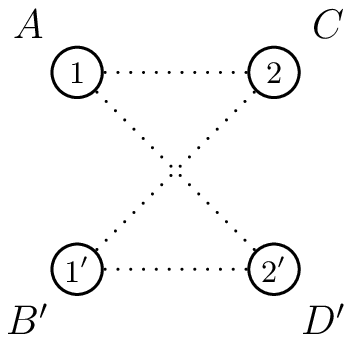}
\vspace{\baselineskip}
\caption{Parties $1,2,1'$ and $2'$ measure observables $A,C,B',D'$ respectively. Dotted lines indicate which parties are jointly considered in the expression \eqref{schumeq1}. Figure taken from Schumacher \cite{schumacher}.}\label{figschum}
\end{figure}

Before assessing this argument let us see why 2-shareability implies the CHSH inequality.  Consider two parties, denoted by  $1$ and $2$. Assume that all possible correlations between parties $1$ and $2$ are 2-shareable to two other parties, denoted  $1'$ and $2'$, that conceivably exist.  Each party has a single system and subjects it to measurement of a single observable, which we denote by $A,C,B',D'$ respectively. See Figure \ref{figschum}. Then for the four possible outcomes of the measurements we get $a(c+d')+b'(c-d')=\pm2$ which implies for the expectation values of the product of the local outcomes
\begin{align}\label{schumeq1}
|\av{AC} +\av{AD'}+\av{B'C}-\av{B'D'}|\leq2.
\end{align}

We now invoke 2-shareability of the correlations to perform the following counterfactual reasoning. By assumption the correlations between parties $1$ and $2$ are the same as between parties $1$ and $2'$. Therefore, if party $2$ would have measured the observable that party $2'$ measured, the observed correlations between the measurements of this observable by party $2$ and the measurement results of party $1$ would be the same as between party $1$ and $2'$. We can thus set $\av{AD'}=\av{AD}$. Analogously we can set $\av{B'C}=\av{BC}$ and $\av{B'D'}=\av{BD}$. Therefore we obtain from \eqref{schumeq1}:
\begin{align}
|\av{AC} +\av{AD}+\av{BC}-\av{BD}|\leq2,
\end{align}
which is the CHSH inequality from which one can prove Bell's theorem. Note that crucial in the argument is that the 2-shareability justifies the counterfactual reasoning.

Although the above argument indeed shows that 2-shareability of correlations already implies a conflict with quantum mechanics, we believe Schumacher's dismissal of issues of locality or local realism in interpreting Bell's theorem to be rather artificial and wanting. 
 
First of all, there is the elementary logical point that, given the violation of the Bell inequality, anything which would imply that it should hold, is false.  Thus, despite Schumacher's argument, it is indeed still the case that quantum mechanics is non-local in the sense that some quantum correlations cannot be given a factorisable form in terms of local correlations, as in \eqref{localdistr}.  

Schumacher was however trying to argue for what \emph{physical message} one should take home from Bell's theorem. Thus although logically Bell's theorem has to do with issues of locality --as was just pointed out--, Schumacher believes the physical message is to be sought elsewhere since he has an alternative (alledgedly) weaker set of assumptions than local realism that imply the CHSH inequality.
 We agree that the weaker the set of assumptions that lead to the Bell-inequality, the more physically relevant the argument becomes\footnote{
 For given the fact that the Bell inequality is violated, we can exclude more and more possible descriptions of nature from using derivations of the Bell inequality that use weaker and weaker assumptions. i.e., more and more possible candidate-theories are then rejected.}.
 
But we question --and here is our second point of critique against Schumacher's dismissal of issues of locality in interpreting Bell's theorem-- whether Schumacher's derivation is indeed logically weaker than standard derivations of Bell's theorem. For all that is needed to get Bell's theorem is the CHSH inequality, and in order to get this from the requirements of the doctrine of local realism we only need to assume that local realism holds \emph{just} for measurement of four different observables: two for party $1$ (e.g., $A,A'$) and two for party 2 (e.g., $B,B'$). Only with respect to these two parties and these four observables we need to assume the correlations to be of the local form \eqref{localdistr}.  It is thus not necessary to assume full blown local realism for an \emph{unlimited} number of observables and parties. 
 
In conclusion, for the purposes of obtaining the CHSH inequality  the assumption of 2-shareability suffices, and so does assuming local realism for measurement of only four observables (two sets of two).  We see no physical reason to believe that the first assumption is weaker than the second. %
 %
   One might object to this reasoning by stating that it is unnatural to require local realism only for measurements between four different observables. For, after all,  we can always think of some extra observables that can be measured over and above the four already specified. But this objection loses its force once one realises that requiring 2-shareability instead of $\infty$-shareability appears to be just as unnatural as requiring local realism for only four observables and two parties instead of for an unlimited number of observables and parties.  For, after all, we can always think of some extra party over and above the two already considered.  Furthermore, we believe it to be telling that in the limit  of an unlimited number of parties $\infty$-shareability and locality are equivalent properties\footnote{Note that locality here means that the correlation is of the form \eqref{localdistr}, without further qualification of the number of parties or the number of possible measurements per party.}, cf. section \ref{sharecorr}.

\section{Monogamy of non-local quantum correlations vs.  monogamy of entanglement}\label{comparingmonogamies}

Two types of monogamy and shareability have been discussed: of entanglement and of correlations (in sections \ref{shst} and \ref{correlationsection} respectively). These are different in principle, although sometimes they go hand in hand. Monogamy (shareability) of entanglement is a property of a quantum state, whereas monogamy (shareability) of correlations
 is not solely determined by the state of the system under consideration, but it is also dependent on 
  the specific setup used to determine the correlations. That is, in the later case it is crucial to also know the number of observables per party and the number of outcomes per observable.  It is thus possible that  a quantum state can give non-local correlations that are monogamous when obtained in one setup, but which are shareable when obtained in another setup. An example of this will be given below. This example also shows that shareability of non-local quantum correlations and shareability of entanglement are related in a non-trivial way.

Masanes \emph{et al.}\cite{masanes06} already remarked (and as was discussed above) that, if we consider an unlimited number of parties, locality and $\infty$-shareability of bi-partite correlations are identical properties. This is analogous to the fact (also discussed above and also obtained by \cite{masanes06}) that quantum separability and $\infty$-shareability of a quantum state are identical in the case of an unlimited number of parties. But if we consider shareability with respect to only one other party the analogy between locality, separability and shareability breaks down. Instead we will show the following result: Shareability of non-local quantum correlations implies shareability of entanglement of mixed states (that gives rise to the non-locality), but not vice versa. The proof for the positive implication runs as follows. Because by assumption the correlations are shareable they are identical for parties $a$ and $b$ and $a$ and $c$. 
The quantum states  $\rho_{ab}$ and $\rho_{ac}$ for the joint systems $ab$ and $ac$ that are supposed to give rise to these correlations
must therefore also be identical, i.e., they are thus shareable.  Furthermore, because the correlations are non-local, these quantum states must be entangled. They furthermore must be non-pure, i.e., mixed, because entanglement of pure states can not be shared. This concludes the proof. Below we give an example of this and show that the converse implication does not hold. In order to do so we will first discuss methods that allow one to reveal the shareability of non-local correlations. 


In general a bi-partite quantum state can be investigated using different setups that each have a different number of observables per party and outcomes per observable. In each such a setup the monogamy and shareability of the correlations  that are obtainable via measurements on the state can be investigated. This is generally performed via a Bell-type inequality that distinguishes local from non-local correlations for the specific setup used.

Let us first assume the case of two parties that each measure two dichotomous observables. For this case the only relevant local Bell-type inequality is the CHSH inequality for which we have seen that the Toner-Verstraete trade-off \eqref{tonerverstraeteineq} implies that all quantum non-local correlations must be monogamous: it is not possible to have correlations between party $a$ and $b$  of subsystem $ab$ and between $a$ and $c$ of subsystem $ac$ such that both $|\av{\mathcal{B}_{ab}}_{\textrm{qm}}|$ and $|\av{\mathcal{B}_{ac}}_{\textrm{qm}}|$ violate the local bound. 
 
It is tempting to think that those entangled states that show monogamy of non-local quantum correlations will also show monogamy of entanglement. This, however, is not the case. For example, three-party pure entangled states exist whose reduced bi-partite states are identical, entangled and able to violate the CHSH inequality (e.g., the $W$-state $\ket{\psi}=(\ket{001}+\ket{010}+\ket{100})/\sqrt{3}$ has such reduced bi-partite   states). These reduced bi-partite states are mixed and their entanglement is shareable, yet, as shown above, they show monogamy of the non-local correlations obtainable from these states in a setup that has two dichotomous observables per party.  Thus we cannot infer from the monogamy of non-local correlations that quantum states responsible for such correlations have monogamy of entanglement; some of them have shareable mixed state entanglement. Consequently,  the study of the non-locality of correlations in a setup that has two dichotomous observables per party, thereby considering the CHSH inequality, does not allow one to reveal shareability of the entanglement of bi-partite mixed states. 

Nevertheless, it is possible to reveal shareability of entanglement of bi-partite mixed states using a Bell-type inequality. But for that it is necessary that the non-local correlations which are obtained from the state in question are not monogamous, i.e., a setup must be used in which some non-local quantum correlations turn out to be shareable.  We have just seen that the case of two dichotomous observables per party, and thus the CHSH inequality,  was shown not to suffice. 
However, adding one observable per party does suffice. Consider the setup where each of the two parties measures three dichotomic observables, which will be denoted by $A,A',A''$ and $B,B',B''$ respectively. Collins \& Gisin \cite{collinsgisin} have shown that for this setup only one relevant new Bell-type inequality besides the CHSH inequality can be obtained (modulo permutations of observables and outcomes). This inequality reads:  
\begin{align}\av{\mathcal{C}}_{\textrm{local}}:=~&\av{AB} +
\av{A'B} +
\av{A''B} +
\av{AB'} +
\av{A'B'} +
\av{AB''} \nn\\&
-\av{A''B'} 
-\av{A'B''} +\av{A} 
+\av{A'} 
-\av{B} 
-\av{B'} 
\leq4,\label{collgisineq}
\end{align}
where for brevity the subscript `local' is omitted in the expectation values in the middle term. Any local correlation must obey this inequality. Collins \& Gisin \cite{collinsgisin} show that the fully entangled pure three-qubit state \begin{align}
\ket{\phi}= \mu \ket{000} +\sqrt{(1-\mu^2)/2}(\ket{110}+\ket{101})
\end{align} 
gives for some values of $\mu$ bi-partite correlations between party $a$ and $b$ of subsystem $ab$ and between $a$ and $c$ of subsystem $ac$ such that the inequality \eqref{collgisineq}  is violated for both these correlations:  
$\av{\mathcal{C}_{ab}}_{\textrm{qm}}=\av{\mathcal{C}_{ac}}_{\textrm{qm}}>4$. This shows that some of the non-local correlations between party $a$ and $b$ can thus be shared with party $a$ and $c$. 

Since $\ket{\phi}$ is a pure entangled three-qubit state the two-qubit reduced states $\rho_{ab}$ and $\rho_{ac}$
 are mixed. Furthermore, since the state $\ket{\phi}$ is symmetric with respect to qubit $b$ and $c$ these reduced states are identical. They must also be entangled because they violate the two-party inequality \eqref{collgisineq}. Therefore, the two-qubit mixed entangled state $\rho_{ab}$ is shareable to at least one other qubit. 
This shows that the inequality \eqref{collgisineq} is  suitable to reveal shareability of entanglement of mixed states. 

It would be interesting to investigate the multi-partite extension of these results.   A preliminary investigation for $N=3$ was performed by Seevinck \cite{seevmon}. There the monogamy of bi-separable three-partite quantum correlations that violate a three-qubit Bell-type inequality that has two dichotomic measurements per party was investigated. For the specific Bell-type inequality under study it was found that maximal violation by bi-separable three-partite quantum correlations is monogamous. This is to be expected because maximal quantum correlations are obtained from pure state entanglement which is monogamous. However, it was found that the correlations that give non-maximal violations can be shared. This indicates shareability of the non-locality of bi-separable three-partite quantum correlations.

\paragraph{Monogamy of non-local correlations is not universal.}
Both the above results by Collins \& Gisin \cite{collinsgisin} and by Seevinck \cite{seevmon} indicate that non-locality of correlations can be shared.  However, as we have stated before, Paw\l owski \& Brukner \cite{pawlowskibrukner} have proven a particular multipartite generalisation of the monogamy constraint \eqref{monogamynosignalling} for a large class of linear multi-partite Bell-type inequalities. 
But the monogamy constraints considered  by  Paw\l owski \& Brukner contain as many Bell-type polynomials $\mathcal{B}$, $\mathcal{C}$, etc. (or Bell-operators in the quantum case) as there are different settings for the different parties. 
The monogamy constraint  $|\av{\mathcal{C}_{ab}}| +|\av{\mathcal{C}_{ac}}| \leq 4$  used above is indeed not of this form, since in the Collins-Gisin inequality the parties each choose between three different settings, not two. The same applies to the Seevinck \cite{seevmon} results.

A closer look at this allows us to phrase a surprising open problem.  This problem reads as follows\footnote{Where it is understood that the Bell-type polynomials $\mathcal{C}$  in Eqns. \eqref{cg1}, \eqref{cg2}, \eqref{pb} are written in a particular way so as to comply with the analysis of 
Paw\l owski \& Brukner \cite{pawlowskibrukner} (i.e. without negative coefficients; every inequality can be brought to that form).}. Consider four parties $a,~b,~c$ and $d$. Collins \& Gisin \cite{collinsgisin} have shown that quantum states for parties $a,b,c$ exist such that
\begin{align}
|\av{\mathcal{C}_{ab}}_{\textrm{qm}}|+ |\av{\mathcal{C}_{ac}}_{\textrm{qm}}|>2LR.
\label{cg1}
\end{align}
Here $LR$ is the bound on $|\av{\mathcal{C}_{ab}}_{\textrm{local}}|$ attainable by local correlations \eqref{localdistr}. By symmetry 
\begin{align}
|\av{\mathcal{C}_{ab}}_{\textrm{qm}}|+|\av{\mathcal{C}_{ad}}_{\textrm{qm}}|>2LR
\label{cg2}
\end{align}
is of course also possible for some quantum states between parties $a,b,d$. However, Paw\l owski \& Brukner \cite{pawlowskibrukner} have derived the monogamy constraint
\begin{align}
|\av{\mathcal{C}_{ab}}_{\textrm{ns}}|+|\av{\mathcal{C}_{ac}}_{\textrm{ns}}|+|\av{\mathcal{C}_{ad}}_{\textrm{ns}}|\leq 3LR
\label{pb}
\end{align}
that must be obeyed by all no-signalling correlations between the four parties $a, \ldots, d$.  Note that \eqref{pb} must also be true for quantum correlations as these are no-signalling. What is not known is whether both \eqref{cg1} and \eqref{cg2} can hold simultaneously for quantum or no-signalling correlations between the four parties $a, \ldots, d$ given the fact that \eqref{pb} must be satisfied by these correlations. It is conjectured\footnote{M.Paw\l owski, private communication.} that this is indeed possible for no-signalling correlations. 

\section{Some consequences for cryptography and quantum key distribution}\label{crypto}

A fruitful application of the monogamy of quantum entanglement is that it provides a basic framework for quantum key distribution.  The reason for this is that entanglement can be seen as the quantum equivalent of what is meant by privacy \cite{horodeckis}.  The main resource for privacy is a secret cryptographic key: correlations shared by interested persons but not known by any other person. Both these two fundamental features of privacy can be found in entanglement: If systems are in a pure entangled state then at the same time (i) the systems are correlated and (ii) no other system is correlated with them, neither quantum mechanically nor classically \cite{koashi}. 

However, this only holds for pure states because we have seen that entanglement of mixed states can be shared.  An example was the $W$-state $\ket{\psi}=(\ket{001}+\ket{010}+\ket{100})/\sqrt{3}$. But we have also seen that in a bi-partite setup with two dichotomous observables per party the non-locality these states can give rise to (i.e., the violation of the CHSH inequality) is monogamous in any no-signalling theory despite the shareability of the entanglement responsible for the non-locality. Quantum key distribution exploits precisely this fact \cite{barrett_key,acin_key,barrettkentpironio,masanes_etal_key,masanes_key}, namely that cryptographic protocols exist where non-locality cannot be shared (i.e., it is monogamous) according to the laws of some general class of theories, namely no-signalling theories. The idea exploited is that a secret key can be generated from correlations that violate the CHSH inequality by a sufficient amount such that the key is secure against eavesdroppers that are only bound by the no-signalling principle.
%
Its basic features can be easily shown in Figure \ref{figmonogamy}. 

Firstly, consider the point $(2,2)$ in this figure. Suppose we force $a$ and $b$ to be maximally classically correlated, which implies that they are perfectly (anti-) correlated (i.e., deterministically) and thus that $\av{\mathcal{B}_{ab}}=2$. Then in any no-signalling theory (including quantum mechanics) this prevents $a$ and $c$  from violating the CHSH inequalities: it must be that $\av{\mathcal{B}_{ac}}\leq 2$ (and analogous for parties $b$ and $c$).

Thus if $a$ and $b$ are perfectly classically correlated, then $b$ and $c$ can share an arbitrary entangled state that is consistent with $a$ and $b$ being perfectly classically correlated, but they will still not be able to violate the CHSH inequality (when $b$ chooses his observables to be the ones in which he is perfectly classically correlated to $a$).

Secondly, consider the points $(2\sqrt{2},0)$ and $(4,0)$ respectively.  This shows that forcing $a$ and $b$ to be maximally non-local in accordance with respectively quantum mechanics or any no-signalling theory forces $a$ and $c$ to have no correlations at all since $\av{\mathcal{B}_{ac}} =0$ in both cases.

Thirdly, consider the region outside the local square with edge points $(\pm2,\pm2)$ but inside the quantum circle. From this we see that as soon as there are non-local correlations between, say, $a$ and $b$ the correlations between $a$  and $c$ cannot be classically maximal, i.e., can not be perfectly (anti-) correlated.  Indeed, a corollary of the quantum monogamy inequality (choosing $C'=C$ in \eqref{tonerverstraeteineq}) gives \cite{toner2}
\begin{align}\av{\mathcal{B}_{ab}}_{\textrm{qm}}^2 + 4\av{AC}_{\textrm{qm}}^2 \leq 8.
\end{align} 
 Thus the stronger $ab$ violate the CHSH inequality, the weaker the correlations of $c$ with $a$; and if $\av{\mathcal{B}_{ab}}=2\sqrt{2}$  then $\av{AC}=0$. Note that a similar result holds for the no-signalling monogamy inequality \eqref{monogamynosignalling} and the points in the region outside the local square but inside the no-signalling tilted square.

Based on these monogamy properties of non-local correlations as well as on the no-signalling  assumption it has been shown recently that quantum key distribution can be secure according to the strongest notion, the so-called universally-composable security \cite{masanes_etal_key,masanes_key}. However, it is noteworthy that Paw\l owski \cite{pawlowski} recently gave a security proof for quantum key distribution protocols based only on the monogamy of non-local correlations, which is a strictly weaker assumption than the assumption of no-signalling.
\\\\
It is important to realise, and we have not seen this stressed anywhere before, that the possibility of non-locality as a resource for secure key distribution depends crucially on the experimental setup (i.e., the number of observables and outcomes) as well as the Bell-type inequality that is being considered. In the previous section we have indeed seen that non-locality is not universally monogamous.
\forget{
Only for the case of two dichotomous observables per party and thus for the CHSH inequality it has been shown that non-locality is monogamous.  For it is the case that just as the monogamy of entanglement does not hold for dimensions of the Hilbert space $d>2$ (a result by Ou \cite{ou}), so does the monogamy of non-local correlations not hold for a situation where we consider correlations between three or more dichotomous measurements per party, instead of just two. Indeed, we have seen in the previous section that for the case of three dichotomous settings per party the non-locality indicated by the Collins-Gisin inequality \eqref{collgisineq} is not monogamous.
}
Examples have been given of non-locality that  is shareable  and which can thus not be used as a resource for secure key distribution.\footnote{The statement by Toner, p 60 in \cite{toner2} that ``$a$ and $b$ cannot violate a Bell inequality even if they share entangled states, if $b$ has to be perfectly correlated to another party $c$'' must thus be qualified. It holds only for the specific Bell inequalities considered by Toner, which, in fact, are the CHSH inequalities.} Nevertheless, it remains remarkable that situations exists where monogamous correlations can be obtained, not just in quantum mechanics but in any no-signalling theory whatsoever.


\section{Discussion}
\label{discussion}
It has been shown that a fruitful way of studying physical theories is via the question whether the physical states and different kinds of correlations that are possible in each theory can be shared to different parties.  Here one focuses on subsets of the parties and whether their states or correlations can be extended to parties not in the original subsets. We have shown that unrestricted general correlations can be shared to any number of parties (called $\infty$-shareable). In the case of no-signalling correlations it was already known that  such correlations can be  $\infty$-shareable  iff the correlations are local. We have shown that this implies that partially-local correlations are also $\infty$-shareable, since they are combinations of local and unrestricted correlations between subsets of the parties. However, it was reviewed that both quantum and no-signalling correlations  that are non-local are not $\infty$-shareable and monogamy constraints for such correlations have also been reviewed.
 
We have investigated the relationship between sharing non-local quantum correlations and sharing mixed entangled states, and already for the simplest bi-partite correlations this was shown to be non-trivial. The Collins-Gisin Bell-type inequality \cite{collinsgisin} indicates that non-local quantum correlations can be shared and this also indicates sharing of entanglement of mixed states. The CHSH inequality was shown not to indicate this. This shows that non-local bi-partite correlations in a setup with two-dichotomous observables per party cannot be shared, whereas this is possible in a setup with one observable per party more. On the quantitative side, we have given a simpler proof of the Toner-Verstraete \cite{tonerverstraete} monogamy relation \eqref{tonerverstraeteineq} as well as a strengthening thereof.  Further, a recently proposed new interpretation of Bell's theorem by Schumacher \cite{schumacher} in terms of shareability of correlations has been critically assessed. Although it is indeed a viable alternative interpretation, we have argued that, contrary to Schumacher's own verdict, it is not weaker, and neither is it more natural than the standard interpretation in terms of the doctrine of local realism.  Finally, we have reviewed the fact  that the monogamy of correlations can be exploited to provide protocols for secure quantum key distribution, and we have indicated that 
some non-local correlations indeed suffice for this task, but that in general not all non-local correlations are monogamous and that this fact should be critically taken into account.

We would like to end by pointing out some open problems and possible avenues for future research. First of all, the relationship between shareability of quantum states and that of non-local quantum correlations asks to be further investigated, thereby extending the analysis of section \ref{comparingmonogamies}. Further, it would be interesting to generalise the monogamy inequality \eqref{tonerverstraeteineq} for quantum correlations from three to $N$ parties.  Also, solving the open problem that was given at the end of section \ref{comparingmonogamies} might reveal fruitful new insight. Lastly, Paw\l owski's investigation and preliminary results  \cite{pawlowski} to base quantum cryptography on the monogamy of correlations only, seems very promising and deserves to be further studied. For this purpose it is desirable to be able to indicate precisely under what conditions non-locality is monogamous, and in precisely what way.
\begin{acknowledgements}
I am grateful to Jos Uffink and Marcin Paw\l owski for fruitful discussions.
\end{acknowledgements}
 \end{document}